\documentclass[a4paper,11pt]{article}
\pdfoutput=1 
\usepackage{graphicx} 
\usepackage{dcolumn} 
\usepackage{bm} 
\usepackage{amssymb} 
\usepackage[latin1]{inputenc}
\usepackage[english]{babel}
\usepackage{shapepar}
\usepackage{jcappub}
\usepackage[T1]{fontenc} 
\usepackage{amsmath}
\usepackage{amsthm}
\usepackage{array}
\usepackage{fancybox}
\usepackage{multirow}
\usepackage{appendix}
\usepackage{epsfig}
\usepackage{amsmath}
\usepackage{amsthm}
\usepackage{array}
\usepackage{fancybox}
\usepackage{multirow}
\usepackage{appendix}
\usepackage{epsfig}
\usepackage{cases}
\usepackage[dvipsnames]{xcolor}
\usepackage{enumitem}
\newcommand{\D}{\partial}
\title{\boldmath Constraining disformal couplings with Redshift Space Distortion}
\author[1]{Avishek Dusoye}
\author[1,2]{, \'Alvaro de la Cruz-Dombriz}
\author[1]{, Peter Dunsby}
\author[3]{and Nelson J. Nunes.}
\affiliation[1]{Cosmology and Gravity Group, Department of Mathematics and Applied Mathematics,\\ University of Cape Town, Rondebosch 7701, Cape Town, South Africa.}

\affiliation[2]{Universidad de Salamanca, Departamento de F\'isica Fundamental, P. de la Merced, Salamanca, Spain}

\affiliation[3]{Instituto de Astrof\'isica e Ci\^encias do Espa\c{c}o,
Faculdade de Ci\^encias da Universidade de Lisboa, Campo Grande, PT1749-016 
Lisboa, Portugal.}
\emailAdd{avsdus001@myuct.ac.za}
\emailAdd{alvaro.dombriz@usal.es}
\emailAdd{peter.dunsby@uct.ac.za}
\emailAdd{njnunes@fc.ul.pt}
\abstract{We study a quintessence model for which the scalar field is disformally coupled to dark matter. The background mimics the $\Lambda$CDM cosmological evolution and the quintessence potential is not specified.  A disformal effect due to the quintessential mass is seen in the growth rate of the cosmological structure on large scales. The disformal parameter renders no appreciable effect on the evolution of the total matter perturbation. An analysis of the conformal parameter and quintessential mass is investigated using the Redshift Space Distortion data to find the best-fit values that might explain the well-known $\sigma_8$ tension.}

\begin{document}
\maketitle

\section{Introduction}
In this era of precision cosmology, the parameters of the $\Lambda$CDM model have been constrained with great accuracy, due to a wide range of cosmological probes, including Planck's latest results\cite{Plank2018, Hildebrandt2016, Heymans2013}. Together with all the compelling evidence for the existence of dark energy \cite{Motta2021,Abbott2017} and dark matter \cite{Arbey2021}, $\Lambda$CDM remains one of the most successful cosmological models from an observational perspective \cite{Peacock1999,Peter2013}. Nevertheless, the $\Lambda$CDM model is unable to explain two phenomena - the unknown nature of the cosmological constant, which is responsible for the accelerated cosmological expansion \cite{Riess1998,Weinberg2012}, and the well-known $\sigma_8$ tension \cite{Verde2019}. This tension  arises because the constraints on clustering, which is imposed by the Planck CMB experiment\cite{Plank2018}, diverges at $2.5\sigma$ confidence level from the large-scale measurement by the Dark Energy Surveys (DES)\cite{Joudaki2019,Riess2016}. This tension, therefore,  may subtly point to the fact that the $\Lambda$CDM model might require some theoretical refinement. Such discrepancy has been a motivation for many attempts to propose several viable theories \cite{DiValentino2021}, which could resolve the issue. Extensions of the $\Lambda$CDM model can be done by modifying either the cosmological fluids or the Einstein-Hibert action. Preserving the Lovelock theorem \cite{Lovelock1971, Lovelock1972}, such a new action can only be achieved by either considering higher dimensions or non-locality or extra fields \cite{Clifton2012}. The potential energy of a scalar field can produce the accelerated cosmological expansion, only if it is light enough to change slowly during the Hubble time \cite{Amendola1999}. The idea of coupled quintessence \cite{Amendola1999, Holden1999} implies that the scalar field or the quintessence is interacting with another fluid while it evolves with cosmic time. Although the nature of the dark components are not known yet, the possibility of their interaction has not been ruled out \cite{Wang2016,Landim2016}. In this article, we firstly assume a quintessence coupled to dark matter and investigate how this coupling influences the evolution of total matter perturbation, as well as the amount of clustering of galaxies. 
The second assumption deals with the geometry in which this quintessence, $\phi$, exists. Let, $g_{ab}$ ,  be the metric of the gravitational geometry, and $\tilde{g}_{ab}$ as the metric of the physical geometry, then both metrics are related to each other by the disformal transformation \cite{Bekenstein1992}:
\begin{equation}\label{E1.1}
\tilde{g}_{ab} = C(\phi)g_{ab} + D(\phi)\D_{a}\phi\D_{b}\phi\;\;,
\end{equation}
where $ C(\phi)$ and $D(\phi)$ are the conformal and disformal functions respectively. In general relativity, $C(\phi)=1$ and $D(\phi)=0$. While the effect of the conformal transformation is to rescale the length of original metric $g_{ab}$ \cite{Clifton2012}, the effect of a disformal transformation can be understood as the distortion in angles and lengths of the original metric  $g_{ab}$ due to compression, which occurs in the direction of the gradient of the field \cite{Bekenstein1992,Teixeira2019}. The disformal framework has gained attention in the last decade \cite{Brax2020, Teixeira2019, Brax2019, Xiao2018, Gannouji2018, Landim2016} and has been applied to many theories such as Horndeski-type scalar tensor theory \cite{Bettoni2013, Zuma2013}, non-linear massive gravity theory \cite{deRham2010,Do2016} and brane cosmology with higher dimensions \cite{Koivisto2013}. Given this promising avenue to explore,  we also assume that the couplings of this coupled quintessence model are of the disformal nature.\\

One could also construct a coupled quintessence model mimicking the $\Lambda$CDM background, so that such a model could both reproduce the successful observational outcome of $\Lambda$CDM, and yet have further degrees of freedom to attempt to alleviate the aforementioned anomalies. For instance, a conformally coupled quintessence (CCQ) with a $\Lambda$CDM background was investigated \cite{Barros2018} in an attempt to alleviate the $\sigma_{8}$ tension. The spherical collapse in coupled quintessence with a $\Lambda$CDM background has been reviewed in Ref. \cite{Barros2019}. The disformal couplings in  $\Lambda$CDM background cosmology was studied in Ref. \cite{Dusoye2020}.\\ 

In this article, we investigate the perturbations for a disformally coupled quintessence (DCQ) model, whose background micmicks $\Lambda$CDM. Thus, this article extends the conformally coupled quintessence of \cite{Barros2018} toward the disformal framework. This article also builds upon the expansion history of Scenario I investigated in \cite{Dusoye2020}.  The effect of quintessential mass and the disformal parameter on the evolution of perturbations is now investigated. Furthermore, an analysis of the disformal parameters, and the quintessential mass are investigated using the Redshift Space Distortion (RSD) data  \cite{delaTorre2013, Okada2015,Guzzo2008,Chuang2013, Blake2012,SDSS2006, GilMarin2015,Tojeiro2012, BOSS2016,Samushia2012,Blake2011, 2dFGRS2004,Howlett2014,Beutler2012,Song2008} to find the best-fit cases of the quintessence model that might explain the well-known $\sigma_8$ tension.\\

The organization of the article is the following. Section 1 provides the motivation of investigating the perturbation theory for a DCQ which is mimicking the $\Lambda$CDM. Section 2 reviews the background cosmology mimicking $\Lambda$CDM. Section 3 revisits  the perturbation equations for the studied DCQ model. Section 4 presents the results and the comparisons with observations.
\section{Background Cosmology mimicking $\Lambda$CDM}
When the Einstein-Hilbert action is extended to the coupled quintessence model, the new action in the Einstein frame is described as: 
\begin{eqnarray} \label{E2.1}
S =  \int \! \left( \dfrac{1}{2 \kappa^2}R \; +  \;\mathcal{L}_{\phi}(g_{ab} ,\phi)\;  + \;\mathcal{L}_{m}(\tilde{g}_{ab} ,\psi)  \right) \,\sqrt{-g} \;\rm{d}^{4}x \;, \\ \nonumber
 \hspace{0.25cm} \text{with}\hspace{0.25cm}  \mathcal{L}_{\phi} = -\dfrac{1}{2}\it  g^{ab}\partial_{a}\phi\partial_{b}\phi - V(\phi).
\end{eqnarray} 
The first term is the standard GR gravitational action, with Ricci scalar $R$ and  $\kappa^2 \equiv 8\pi G = M_{\rm{pl}}^{-2}$. The Planck mass is taken to be $ M_{\rm{pl}} = 2.4 \times 10^{18} \; \rm{GeV}$ with the speed of light and reduced Planck constant set to unity $(\hbar = c  = 1)$. $\mathcal{L}_{\phi}$ is the Lagrangian for the quintessence and consists of a kinetic term and a potential function $ V(\phi)$. $\mathcal{L}_{m}$ is the Lagrangian for matter. The metric variation of \eqref{E2.1} with respect to the metric $g^{ab}$ results in the field equations in the Einstein frame.
\begin{equation} \label{E2.2}
G_{ab} \equiv R_{ab} - \dfrac{1}{2}g_{ab}R = \kappa^2 T_{ab}^{(tot)} =\;   \kappa^2 ( T_{ab}^{\phi} + T_{ab}) \;,
\end{equation}
where $G_{ab}$, $R_{ab}$, $T_{ab}^{\phi}$, and $T_{ab}$ are the Einstein tensor, the Ricci tensor and the energy-momentum tensors for quintessence and matter respectively. The variation of the action \eqref{E2.1} with respect to the quintessence $\phi$ results in the modified Klein-Gordon equation \cite{Carsten2016a}.
\begin{equation}\label{E2.3}
\square\phi = V_{,\phi} - Q_0 \;,
\end{equation}
where $\square \equiv g^{ab}\nabla_{a}\nabla_{b} $ is the D'Alembertian operator and $V_{,\phi}$ denotes the derivative of $V(\phi)$ with respect to $\phi$. Let "$\rm{f}$" denote any general coupled fluid. The conservation equations for the energy-momentum tensors $T_{ab}^{\rm{f}}$ and $T_{ab}^{\phi}$ in the coupled quintessence model are given below:
\begin{eqnarray} \label{E2.4}
\nabla^{a}T_{ab}^{\phi}= -Q_0\D _{b} \phi  \;,  \hspace{0.5cm} \text{and} \hspace{0.5cm}
\nabla^{a}T_{ab}^{\rm{f}}= +Q_0 \D _{b} \phi    \;,                                                          
\end{eqnarray} 
so that there is a flow of energy and momentum between two coupled fluids in the model, which is governed by the interaction term $Q_0$. The total energy-momentum tensor is conserved although their individual components are not separately conserved. For the  disformal framework, this  interaction term for the coupling between the quintessence and a single fluid can be determined \cite{Carsten2016a}.
\begin{equation}\label{E2.5}
Q_0 = \dfrac{C_{,\phi}}{2C}T + \dfrac{D_{,\phi}}{2C}T^{ab}\nabla_{a}\phi\nabla_{b}\phi - \nabla_{a}\left( \dfrac{D}{C}T^{ab} \nabla_{b}\phi \right) \;,
\end{equation}
where $C_{,\phi}$ and $D_{,\phi}$ are the  derivatives of conformal and disformal functions $C(\phi)$ and $D(\phi)$ respectively appearing in the disformal transformation \eqref{E1.1}. The spatially flat Friedmann-Lema\^{i}tre-Robertson-Walker (FLRW) is chosen to be the background metric, where $a(t)$ is the scale factor, $t$ is the cosmic time and $i,j= 1,2,3$ denote the spatial coordinates.
\begin{equation}  \label{E2.6}
ds^2 = g_{ab}dx^{a}dx^{b} = -dt^2 + a^2(t)\;\delta_{\textit{ij}}dx^{\textit{i}}dx^{\textit{j}} \;,
\end{equation}
The Friedmann and Raychaudhuri equation for the coupled quintessence can then be obtained \citep{Peter2013}, where the dot denotes a derivative with respect to cosmic time:
\vspace{-0.2cm}
\begin{equation} \label{E2.7}
  H^{2}  = \dfrac{\kappa^2}{3}(\rho_{\phi} + \rho_{\rm{f}})\;, ~~
  \dot{H} = -\dfrac{1}{2}\kappa^2\left[ (\rho_{\phi} + P_{\phi}) + \rho_{\rm{f}}(1+w_{\rm{f}})\right] \;,
\end{equation}
where $H \equiv \dot{a}/a$ is the Hubble parameter. Here $\rho_{\rm{f}}$ and $w_{\rm{f}}$ are the energy density and the equation of state respectively for a general fluid coupled to the quintessence. The energy density and the pressure for the quintessence is denoted by $\rho_{\phi}$ and $P_{\phi}$ and are defined as: 
\begin{equation} \label{E2.8}
\rho_{\phi} \equiv \dfrac{1}{2}\dot{\phi}^{2} + V(\phi) \hspace{0.5cm}\text{and}\hspace{0.5cm} P_{\phi} \equiv \dfrac{1}{2}\dot{\phi}^{2} - V(\phi)\;.
\end{equation}
Assuming a $\Lambda$CDM background, $\rho_c$ and $\rho_{\Lambda}$ denote the energy density of the cold dark matter and the dark energy in the form of a cosmological constant respectively. When the quintessence is coupled to cold dark matter i.e, $\rho_{\rm{f}} = \rho_c$, one can deduce the following\cite{Barros2018, Dusoye2020}:
\begin{eqnarray} \label{E2.9}
H^{2} =   \dfrac{\kappa^2}{3}(\rho_{\phi} + \rho_{\rm{f}}) =  \dfrac{\kappa^2}{3}(\rho_{\Lambda} + \rho_{c})  = H^{2}_{\Lambda \rm{CDM}}  \hspace{0.25cm} &\Rightarrow&  \hspace{0.25cm}\rho_{\phi} = \rho_{\Lambda} \;,\\ \label{E2.10}
 \hspace{-0.3cm}\dot{H} = -\dfrac{1}{2}\kappa^2\left[ (\rho_{\phi} + P_{\phi}) + \rho_{\rm{f}}(1+w_{\rm{f}})\right]  =   -\dfrac{1}{2}\kappa^2\left[ (\rho_{\Lambda} + P_{\Lambda}) + \rho_c\right]   =   \dot{H}_{\Lambda \rm{CDM}} \hspace{0.1cm}&\Rightarrow & \hspace{0.1cm}P_{\phi} = -\rho_{\Lambda},\hspace{1.0cm} \\ \label{E2.11}
\Rightarrow \hspace{0.25cm} V(\phi)=\dfrac{1}{2}\dot{\phi}^2 + \rho_{\Lambda}  \hspace{0.25cm} \text{and} \hspace{0.25cm}\  V_{,\phi} = \ddot{\phi}\;.
\end{eqnarray} 
Since  $V(\phi)$ can be expressed in terms of $\phi$ and $\rho_{\Lambda}$ via Equation \eqref{E2.11} one does not have to specify the scalar potential. By projecting the conservation equations \eqref{E2.4} along the velocity of a co-moving observer $u_{a}$, we obtain the continuity equations for each fluid:
\begin{eqnarray}
 \ddot{\phi} +3H\dot{\phi} + V_{,\phi} &=& +Q_0 \;\;,                  \label{E2.12} \\
 \dot{\rho}_{\rm{c}} + 3H\rho_{\rm{c}}&=& -Q_0\dot{\phi} \;.    \label{E2.13}
 \end{eqnarray}
 \subsection{Construction of the Dynamical System}
The dynamical system of this cosmological model can be constructed by  defining appropriate dimensionless dynamical variables \cite{Carsten2016a}. The dynamical variable $\sigma$ measures the strength of the disformal coupling. 
\begin{eqnarray}  \label{EDV}
x^2 \equiv  \dfrac{\kappa^2\dot{\phi}^{2} }{3H^2} , \hspace{0.5cm}
y^2 \equiv  \dfrac{\kappa^2\rho_{\Lambda}}{3H^2} ,\hspace{0.5cm}
z^2 \equiv \dfrac{\kappa^2\rho_{c}}{3H^2}  ,\hspace{0.5cm} 
\sigma \equiv  \dfrac{D(\phi)}{\kappa^2 C(\phi)}H^2 ,\hspace{0.5cm} \nonumber \\
\lambda_C \equiv - \dfrac{C_{,\phi}}{\kappa C} ,\hspace{0.5cm}
\lambda_D \equiv - \dfrac{D_{,\phi}}{\kappa D}\;,\hspace{0.5cm}
\Lambda_C \equiv  \dfrac{C_{,\phi\phi}}{\kappa^2 C} ,\hspace{0.5cm}
\Lambda_D \equiv  \dfrac{D_{,\phi\phi}}{\kappa^2 D}\;.\hspace{0.5cm}
\end{eqnarray}
The variables $\Lambda_C$ and $\Lambda_D$ are defined for the cases where $\lambda_C$ and $\lambda_D$ might be field dependent (See Equation \eqref{E2.20}). When the continuity equations \eqref{E2.12} and \eqref{E2.13} are expressed in terms of the dynamical variables, one can obtain the set of evolution equations in terms of e-fold $N=\ln a$ \cite{Dusoye2020}:
\vspace{-0.5cm}
\begin{eqnarray}  \label{E2.15}
 x^{\prime} &=& -\dfrac{H^{\prime}}{H}x-\dfrac{3x}{2} +\dfrac{\sqrt{3}\tilde{Q}_0 z^2}{2} \;,  
\\  \label{E2.16}
 z^{\prime} &=& -\dfrac{H^{\prime}}{H}z - \dfrac{3z}{2} -\dfrac{\sqrt{3}\tilde{Q}_0}{2}xz\;, 
 \\ \label{E2.17}
 \sigma^{\prime} &=& \left[\sqrt{3} x \left(\lambda _C-\lambda _D\right)-3 \left(x^2+z^2\right)\right] \sigma\;,
 \\ \label{EQ0}
\tilde{Q}_0 \equiv  \dfrac{\kappa Q_0}{3H^2 z^2} &=& \frac{\lambda _C \left(1-6 \sigma  x^2\right)+3 \sigma  x \left(x \lambda _D+\sqrt{3}\right)}{2 -6 \sigma x^2 + 3 \sigma z^2}\;,
 \end{eqnarray}
where $^{\prime}$ denotes derivative with respect to $N$. These variables $\eqref{EDV}$ are constrained to obey the Friedmann and Raychaudhuri equations \eqref{E2.7}, which now become:
 \begin{eqnarray}\label{E2.19}
1= x^2 + y^2 + z^2\;, \hspace{0.5cm} \text{and} \hspace{0.5cm} \hat{H} \equiv \dfrac{H^{\prime}}{H} =-\dfrac{3}{2}\left(x^2 + z^2\right) \hspace{0.5cm} \text{and} \hspace{0.5cm} \dfrac{H^{\prime\prime}}{H} =6(x^2 + z^2) \;.
\end{eqnarray}
The derivative of the Raychaudhuri equation is also computed and will be useful later in Section 3.3. The derivative of the variables $\lambda _C$ and $\lambda _D$ with respect to $N$ are given below:
\begin{eqnarray} \label{E2.20} 
\lambda _C^{\prime} = \sqrt{3}x(\lambda _C^2-\Lambda _C)\hspace{0.5cm} \text{and}\hspace{0.5cm}\lambda _D^{\prime} = \sqrt{3}x(\lambda _D^2-\Lambda _D)\;. 
\end{eqnarray} 
Finally, in order to close this system of differential equations, we specify the conformal and disformal coefficients as in the previous literature \cite{Carsten2016a} :
\begin{eqnarray}  \label{E2.21} 
C(\phi) = \exp(2\alpha \kappa \phi)  \;, \hspace{1.0cm}
D(\phi) &=& \dfrac{\exp\left[2(\alpha + \beta) \kappa \phi\right]}{M^4}\;=D_m^{-4}\exp\left[2(\alpha + \beta) \kappa \phi \right]  \\  \label{E2.22}
\lambda_C \equiv - \dfrac{1}{\kappa}\dfrac{C_{,\phi}}{C} =-2\alpha \;, \hspace{0.75cm}
\lambda_D &\equiv& - \dfrac{1}{\kappa}\dfrac{D_{,\phi}}{D} =-2(\alpha + \beta)  \;, 
\end{eqnarray} 
where the parameters $\alpha$ and $\beta$ are constants. $M=D_m^{-1}$ is the mass scale in the disformal coupling. For this choice of $C(\phi)$ and $D(\phi)$ which yields $\lambda_C$ and $\lambda_D$ as constants,  one obtains $\lambda _C^{\prime}=\lambda _D^{\prime} = 0$. In a conformal case, the function $D(\phi)$ and its derivatives vanish.
\section{Perturbative Cosmology}
As mentioned above, we shall investigate the perturbations of a disformally coupled quintessence model, whose background mimics $\Lambda$CDM (See Section 2). We therefore study the evolution of the density perturbations for the coupled fluid and the growth factor in this context and compare it to observations. The scalar perturbations in the Newtonian gauge (i.e., the Bardeen potentials $\Phi$ and $\Psi$)  are considered along the FLRW metric, whose line element is given:
\begin{equation}  \label{E3.1}
ds^2 = -(1+2\Phi)dt^2 + a^2(t)\;\delta_{\textit{ij}}(1-2\Psi)dx^{\textit{i}}dx^{\textit{j}} \;.
\end{equation}
Similarly, the energy density $\rho_n$ and the pressure $P_n$ of the $n^{th}$ fluid as well as the scalar field  $\phi$ have been perturbed in order to compute the perturbed energy-momentum tensors:
\begin{eqnarray}  \label{E3.2} 
\rho_n(\vec{x},t) &= &\bar{\rho}_n(t) + \delta\rho_n(\vec{x},t)\;,\\
P_n(\vec{x},t) &=& \bar{P}_n(t) + \delta P_n(\vec{x},t) \nonumber \;,\\ 
\phi(\vec{x},t) &=& \bar{\phi}(t) + \chi (\vec{x},t) \nonumber \;,
\end{eqnarray} 
so that one can define the density contrast $\delta_n \equiv  \delta\rho_n/\bar{\rho}_n$ and sound speed $c_s^2 \equiv  \delta P_n /\delta\rho_n $ of the  $n^{th}$ fluid. The scalar modes of the perturbed Einstein equations (i.e. $\delta G^a_b =  \kappa^2 \delta T^a_b$) for the coupled quintessence for different combinations of $(a,b)$ are given by the following equations in cosmic time:
\vspace{-0.25cm}
\begin{eqnarray} \label{E3.3} 
\nabla^2(\dot{\Psi} + H\Psi) &=& \dfrac{\kappa^2 \dot{\phi}}{2}\nabla^2\chi -\dfrac{3}{2}a H^2(\Omega_{c}\theta_{c} + \Omega_{\mathtt{B}}\theta_{\mathtt{B}})\;, \\ \label{E3.4} 
a^{-2}\nabla^2\Psi - 3 H(\dot{\Psi} +  H\Psi) &=& \dfrac{3}{2} H^2(\Omega_{c}\delta_{c} + \Omega_{\mathtt{B}}\delta_{\mathtt{B}}) +  \dfrac{ \kappa^2}{2}(\dot{\phi}\dot{\chi}-\Psi\dot{\phi}^2 + V_{,\phi}\chi)\;, \\ \label{E3.5}
\ddot{\Psi} + 4H\dot{\Psi} + (2\dot{H} + 3H^2)\Psi  &=& \dfrac{3}{2} H^2 c_s^2(\Omega_{c}\delta_{c} + \Omega_{\mathtt{B}}\delta_{\mathtt{B}}) +  \dfrac{\kappa^2}{2}(\dot{\phi}\dot{\chi}-\Psi\dot{\phi}^2 - V_{,\phi}\chi)\;,
\end{eqnarray}
where the equations \eqref{E3.3}-\eqref{E3.5} make use of the relation  $\Psi = \Phi$, which has been obtained from the traceless spatial components of the perturbed Einstein equations. The variable $\theta_{n} \equiv \nabla^2v_{n}$ is the perturbed co-moving velocity  and $\Omega_n = \frac{\kappa\rho_n}{3H^2}$ is the background density parameter for the n$^{th}$ fluid. The subscript $``\mathtt{B}"$ denotes baryons. The perturbed Klein-Gordon equation for the quintessence and the perturbed conservation and continuity equations for any general coupled fluid are given below \citep{VandeBruck2012X}:
\begin{eqnarray} 
\label{E3.6} 
\ddot{\chi}  &=&- 3H\dot{\chi} + a^{-2}\nabla^2\chi - V_{,\phi\phi}\chi + \dot{\phi}(\dot{\Phi} + 3\dot{\Psi}) + \frac{Q_0\dot{\chi}}{\dot{\phi}} + Q_1 + (2\ddot{\phi} + 6H\dot{\phi})\Phi \;,
\end{eqnarray}
\begin{eqnarray} 
\label{E3.7} 
\dot{\delta}_{\rm{f}}   &=& - (1+w_{\rm{f}})\left(\dfrac{\theta _{\rm{f}}}{a} -3\dot{\Psi}\right) - 3H (\textit{c}_s^2-w_{\rm{f}})\delta_{\rm{f}}  + \dfrac{Q_0\,\dot{\phi}}{\rho_{\rm{f}} }\delta _{\rm{f}} - \dfrac{Q_0\,\dot{\chi}}{\rho_{\rm{f}}} -  \dfrac{Q_1\,\dot{\phi}}{\rho_{\rm{f}} }\;,\\
\label{E3.8} 
\dot{\theta}_{\rm{f}}    &=& - H(1-3w_{\rm{f}})\theta _{\rm{f}} - \dfrac{\dot{w_{\rm{f}}}}{1+w_{\rm{f}}}\theta _{\rm{f}} - \dfrac{\nabla^2\Psi}{a} - \dfrac{\textit{c}_s^2 \nabla^2\delta _{\rm{f}}}{(1+w_{\rm{f}})a} + \dfrac{w_{\rm{f}}\nabla^2\pi_{\rm{f}}}{(1+w_{\rm{f}})}+ \dfrac{Q_0\,\dot{\phi}}{\rho_{\rm{f}} }\theta _{\rm{f}} + \dfrac{Q_0 \nabla^2\chi}{(1+w_{\rm{f}})\rho_{\rm{f}}  a}\;,\hspace{1.0cm}
\end{eqnarray}
where $V_{,\phi\phi}$ is the second derivative of the potential $V(\phi)$ with respect to quintessence. The equations \eqref{E3.7} and   \eqref{E3.8} also utilized the relation $\Psi = \Phi$. The variable $\pi_{\rm{f}}$ is the scalar part of the anisotropic shear perturbation $\pi^{i}_{j}$ from the spatial component of the perturbed energy momentum tensor i.e. $T^{i}_{j} = (\bar{P} + \delta P)\delta^{i}_{j}  + \pi^{i}_{j}$. The variable $\pi_{\rm{f}}$ is often related to shear stress $\sigma_{\rm{f}}$ by $\sigma_{\rm{f}} = 2\pi_{\rm{f}}w_{\rm{f}}/3(1 + w_{\rm{f}})$ in the literature \cite{MaBertschinger1995}. The term $Q_1$ is the perturbation of the interaction term $Q_0$ from \eqref{E2.5}  and is defined in cosmic time as follows \citep{VandeBruck2012X}:
\begin{eqnarray}
\label{E3.9} 
Q_1 &=& \dfrac{-\rho_{\rm{f}} }{C+D(\rho_{\rm{f}}  -\dot{\phi}^2)}\left(\mathcal{B}_1\delta_{\rm{f}}  + \mathcal{B}_2\dot{\Phi}   +\mathcal{B}_3\Psi +  \mathcal{B}_4\dot{\chi} + \mathcal{B}_5\chi \right)\;,\\
\nonumber \\ 
\label{E3.10} 
\text{with} \hspace{0.5cm} 
\mathcal{B}_1 &=& \dfrac{C_{,\phi}}{2}(1-3\textit{c}_s^2) - 3DH\dot{\phi}(1+ \textit{c}_s^2) - D(V_{,\phi}  - Q_0) - D\dot{\phi}^2\left( \dfrac{C_{,\phi}}{C} - \dfrac{D_{,\phi}}{2D}\right)\;, \\
\label{E3.11} 
\mathcal{B}_2 &=& 3D\dot{\phi}(1+w_f)\;,  \\ 
\label{E3.12} 
\mathcal{B}_3 &=& 6DH\dot{\phi}(1+w_{\rm{f}})  + 2D\dot{\phi}^2\left( \dfrac{C_{,\phi}}{C} - \dfrac{D_{,\phi}}{2D} + \dfrac{Q_0}{\rho_{\rm{f}} } \right)\;, \\
 \label{E3.13} 
\mathcal{B}_4 &=& -3DH(1+w_{\rm{f}}) -  2D\dot{\phi}\left( \dfrac{C_{,\phi}}{C} - \dfrac{D_{,\phi}}{2D} + \dfrac{Q_0}{\rho_{\rm{f}} } \right)\;,\\
 \label{E3.14} 
\mathcal{B}_5 &=&  \dfrac{C_{,\phi\phi}}{2}(1-3w_{\rm{f}}) + \dfrac{(1+w_{\rm{f}})D\nabla^2}{a^2} -DV_{,\phi\phi} - D_{,\phi}V_{,\phi} - 3D_{,\phi}H\dot{\phi}(1+w_{\rm{f}})\\ 
&& -D\dot{\phi}^2\left[\dfrac{C_{,\phi\phi}}{C} + \left(\dfrac{C_{,\phi}}{C}\right)^2 + \dfrac{C_{,\phi}D_{,\phi}}{CD} - \dfrac{D_{,\phi\phi}}{2D}\right] + \left(C_{,\phi} + D_{,\phi}\rho_{\rm{f}}  - D_{,\phi}\dot{\phi}^2\right)\dfrac{Q_0}{\rho_{\rm{f}} } \;. \nonumber 
\end{eqnarray}
The $\nabla^2$ operator appearing in the second term of the coefficient $\mathcal{B}_5$ in \eqref{E3.14} acts on the variable $\chi$ in equation \eqref{E3.9}. The perturbed Einstein equations \eqref{E3.3}-\eqref{E3.5}, the perturbed conservation and continuity equations \eqref{E3.6}-\eqref{E3.8} are used in deriving the second order evolution equations of density perturbations. The corresponding evolution equations for the baryonic density perturbations are however much simpler, because they are not coupled to the quintessence i.e., $Q_0 = Q_1 =0 $ for baryons, and therefore The evolution equations of baryonic density perturbations is then obtained,  as shown by  Ref. \citep{Barros2018}.
\begin{eqnarray}\label{E3.15}
\delta_{\mathtt{B}}^{\prime\prime} + \delta_{\mathtt{B}}^{\prime}\left(2+ \hat{H}  \right) - \dfrac{3}{2}\left(\Omega_{\mathtt{B}}\delta_{\mathtt{B}} +  \Omega_{c}\delta_{c}\right) = 0\;.
\end{eqnarray}
In our model, since it is cold dark matter which is coupled to quintessence and it is also assumed as a perfect fluid, one can set the variables $w_{\rm{f}}= \dot{w}_f = \nabla^2\pi_{\rm{f}} = 0$ in the 
equations \eqref{E3.7}-\eqref{E3.8}. Using the equations \eqref{E3.7}-\eqref{E3.8}, the second-order evolution equation for the density perturbations of the coupled DM can be computed after some arrangements, as shown by the corresponding equation (4.24) in Ref. \cite{Gleyzes2015a} (See appendix \ref{A1}).
\begin{eqnarray}\label{E3.16}
&& \delta_c^{\prime\prime}  \;+\; \delta_c^{\prime}\left(2-2\sqrt{3}\,\tilde{Q}_0\,x + \hat{H}  \right) - \delta_c\left[\sqrt{3}\,\hat{q}_0x  + \frac{\tilde{Q}_0}{2}\left(7\sqrt{3}x + 3\tilde{Q}_0z^2\right) \right]   \nonumber \\
&& +\; \sqrt{3}\,\hat{q}_1\textit{x}\; +\;\frac{\tilde{Q}_1}{2}\left(7\sqrt{3}x-3\tilde{Q}_0z^2\right) -\dfrac{3}{2}(z^2\delta_{c} + \Omega_{\mathtt{B}}\delta_{\mathtt{B}}) \;=\; 0  \;,
\end{eqnarray}
where we make use of the dynamical variables \eqref{EDV} and further introduce the normalised variables, which are defined below:
\begin{eqnarray}  \label{E3.17}
\tilde{Q}_0 \equiv  \dfrac{\kappa Q_0}{3H^2 z^2} ,\hspace{1cm}
\tilde{Q}_1\equiv  \dfrac{\kappa Q_1}{3H^2 z^2} ,\hspace{1cm}
\hat{q}_0 \equiv \dfrac{\kappa Q_0^{\prime}}{3H^2 z^2}  ,\hspace{1cm}
\hat{q}_1 \equiv \dfrac{\kappa Q_1^{\prime}}{3H^2 z^2} \;.
\end{eqnarray}
While the tilde (i.e., in $\tilde{Q}_0$ and $\tilde{Q}_1$) denotes the normalisation of the interaction term $Q_0$  and of its perturbation $Q_1$, the hat (i.e., in $\hat{q}_0$ and $\hat{q}_1$) denotes the normalisation of the derivative of the interaction term $ Q_0^{\prime}$ and the derivative of its perturbation $Q_1^{\prime}$. The normalised derivatives $\hat{q}_0 $ and $\hat{q}_1$, which is required to complete the evolution equation \eqref{E3.17}, can be computed for arbitrary choice of $\lambda_C$ and $\lambda_D$ as follows:
\begin{eqnarray} \label{E3.18}
\hat{q}_0 &=& -\tilde{Q}_0 \left(  3+ \sqrt{3}\tilde{Q}_0x\right) + \left(3\tilde{Q}_0+\dfrac{3}{2}(\lambda_D -2\lambda_C)\right)\hspace{-0.15cm}\left(\dfrac{\sigma^{\prime}x^2 + 2\sigma xx^{\prime}}{1-3\sigma x^2}\right)   \nonumber \\
 &&+\dfrac{\sqrt{3}}{2}\left(\dfrac{\sigma^{\prime}-\hat{H}\sigma }{1-3\sigma x^2}\right)\hspace{-0.2cm}\left(3x-  \sqrt{3}\tilde{Q}_0 z^2 \right)-\sqrt{3}\sigma\hspace{-0.15cm}\left[\dfrac{x^{\prime\prime} + 2\hat{H}x^{\prime} +  6x(x^2+z^2)}{1-3\sigma x^2} \right]\;. \\
\nonumber  \\ 
 \label{E3.19}
\hat{q}_1 &=& \dfrac{3\tilde{Q}_1\left(\sigma^{\prime}x^2 + 2\sigma xx^{\prime} -2\sigma^{\prime}z^2 - 4\sigma zz^{\prime} \right)+\left[ \left( 3+ \sqrt{3}\tilde{Q}_0 x  \right)\delta_{c} -\delta_c^{\prime}\right]\mathcal{A}_1 - \mathcal{A}_1^{\prime}\delta_{c}}{\left(1-3\sigma x^2+ 6\sigma z^2\right)}\;,\hspace{1.2cm}
\end{eqnarray}
such that $\mathcal{A}_1 = \mathcal{B}_1/\kappa C$ as per equation \eqref{E6.5} in appendix \ref{A3} and the term $\mathcal{A}_1^{\prime}$ is given by:
\begin{eqnarray} 
\label{E3.20}
\mathcal{A}_1^{\prime} &=& - \dfrac{\lambda_C^{\prime}}{2} - \sqrt{3}\left(\sigma^{\prime} -  \dfrac{H^{\prime}}{H}\sigma \right)\hspace{-0.2cm}\left(x^{\prime} -  \dfrac{H^{\prime}}{H}x \right) -\sqrt{3}\sigma\left( x^{\prime\prime} + 2 \dfrac{H^{\prime}}{H}x^{\prime} + \dfrac{H^{\prime\prime}}{H}x \right)\nonumber \hspace{1.0cm}\\
 &&-\dfrac{3}{2}\left(\sigma^{\prime}x^2 + 2\sigma xx^{\prime}\right)\left(\lambda_D - 2\lambda_C \right) - \dfrac{3\sigma x^2}{2}\left(\lambda_D^{\prime} - 2\lambda_C^{\prime} \right)\;. 
\end{eqnarray}
The steps to compute the terms $\hat{q}_0$ and $\hat{q}_1$ is detailed in the appendix \ref{A2} and \ref{A3} .
\section{Results}
The disformally coupled quintessence model, which is described by the system of  evolution equations \eqref{E3.15} and \eqref{E3.16}, is solved numerically in order to obtain the perturbations $\delta_c$ and $\delta_b $ in terms of cosmological redshift $Z = 1/a -1 $. Note that the redshift $Z$ is not to be confused with dynamical variable $z$. The perturbation equation \eqref{E3.16} includes the normalised interaction term $\tilde{Q}_0$ in equation \eqref{EQ0}, the derivative $\hat{q}_0$ in equation \eqref{E3.18}, the  normalised perturbed interaction term $\tilde{Q}_1$ in equation \eqref{E6.11}, and the derivative $\hat{q}_1$ in equation \eqref{E3.19}. Since the perturbation equations \eqref{E3.15} and \eqref{E3.16} also depend on background dynamical variables $x$ and $z$, the numerical analysis requires solving the background dynamical system \eqref{E2.15}-\eqref{EQ0} first. From the above definition \eqref{EDV}, one evaluates the initial values of the dynamical variables at redshift $Z_{i}$ for an early universe according to the relations,
\begin{eqnarray}  \label{E4.1}
x_{i}=0 ,\hspace{0.5cm}
y_{i}=\sqrt{\frac{\Omega _{\Lambda 0}}{h^2_{i}}},\hspace{0.5cm}
z_{i}&=&\sqrt{\frac{\Omega _{\rm{cdm}0}(1+Z_{i})^3}{h^2_{i}}},\hspace{0.5cm} 
\sigma_{i} =  \dfrac{D H^2_{0}}{\kappa^2 C}h^2_{i}, \hspace{0.5cm}
h_{i}=h(Z_i),
\end{eqnarray}
where $h^2(Z) \equiv H^2/H_{0}^2$ is the reduced Hubble function and $H_{0}$ is Hubble parameter today. The density parameters $\Omega _{\rm{cdm}0}$ and $ \Omega _{\Lambda 0}$ correspond to DM and DE evaluated today.  The initial values of redshift, density perturbations and their derivatives are set as:
\begin{eqnarray}  \label{E4.2}
 Z_i=10^{4},\hspace{0.5cm}\text{and}\hspace{0.5cm}
 \delta_c(Z_i)\;=\;\delta_b(Z_i)\;=\;10^{-3} ,\hspace{0.5cm}\text{and}\hspace{0.5cm}
 \delta_c^{\prime}\;=\;\delta_b^{\prime}\;=\;10^{-3}.
\end{eqnarray}
 We shall now investigate how the density contrasts will eventually grow or decrease as the cosmological model enters into the non-linear regime for the given set of initial conditions \eqref{E4.1} and \eqref{E4.2}. Observationally, the length scale  $R_8= 8h^{-1}\;\textrm{Mpc}$ is defined as the spherical radius below which the linear regime is not valid. The parameter $\sigma^0_8$ is the present amplitude of the matter power spectrum at the scale of $R_8$  \cite{Peter2013}. Therefore, this disformally coupled quintessence model has four free parameters: $\Theta_{\mu}=(\;\sigma^0_8,\;\alpha,\;\beta$ and $ D_m\;)$.
 
Additionally, in order to ensure that both we recover $\Lambda$CDM at high redshifts in matter-dominated epoch and that we can safely use the power spectrum as predicted by GR, we impose that the parameters $\alpha$ and $\beta$ are set as zero for any redshift $Z> 100$ since they are only allowed to influence the evolution of density perturbations for $Z<100$ until present.
 
\begin{figure}[h!] 
\vspace{-0.2cm}
\center
\includegraphics[scale=0.4]{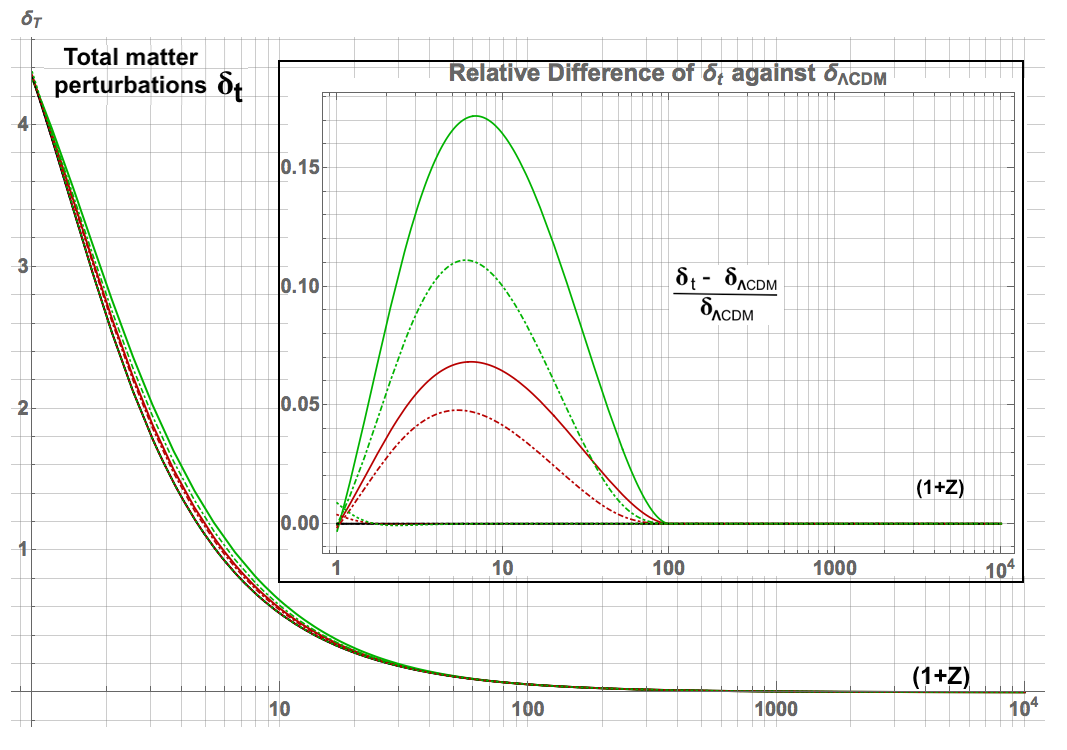} 
\caption{\label{fig1}The effect of quintessential mass upon evolution of total matter perturbations is shown. The black line represents $\Lambda$CDM (i.e. $\alpha = 0.0$ and $D_m = 0.0$) while the red and green solid lines correspond to $\alpha = 0.2$ and $0.32$ respectively. For each color line, the solid, dot-dashed and dotted lines correspond to $D_m = 0.001,\; 0.05,\;1.0\; \rm{meV}^{-1}$ respectively. The dotted lines concides with black line until at low redshift. The parameter $\sigma^0_8=0.818$ from Ref. \cite{Barros2018} and $\beta$ is set as unity.}
\end{figure}

\hspace{-0.8cm}For the analysis, we define (i) the  total matter density perturbations $\delta_t $,  (ii) the growth function g, which describes the density perturbations with respect its current value, and (iii) the growth rate $f$, indicating the rate of evolution for the total density perturbations \cite{Barros2018}:
\begin{eqnarray}  \label{E4.3}
\delta_t \equiv \dfrac{\Omega_b\delta_b  +  z^2\delta_c}{\Omega_b + z^2} , \hspace{0.5cm} g(Z)=\dfrac{\delta_t(Z)}{\delta_0} \hspace{0.5cm}\text{and}&&\hspace{0.5cm} f= \dfrac{d\ln \delta_t }{d\ln a} =\dfrac{\delta_t^{\prime}}{\delta_t}, \hspace{1.5cm} \\   
 \hspace{0.1cm}\text{such that}&& \hspace{0.2cm} f\sigma_8(Z)=fg\sigma^0_8=\left(\dfrac{\delta_t^{\prime}}{\delta_0}\right)\sigma^0_8\;, \hspace{1.0cm}  \nonumber
\end{eqnarray}
where $\rm{f}\sigma_8(Z)$ characterises the amount of clustering of galaxies at given redshift. Thus, after obtaining a numerical solution of $\delta_c$ and $\delta_b$ for a given choice of the parameters of the model, one can plot the numerical $f\sigma_8(Z)$ curve with respect to redshift $Z$. The disformal parameter $\beta$ was found to have no appreciable effect on the evolution of total matter perturbation $\delta_t$ and $f\sigma_8(Z)$. The relative difference by percentage of $\delta_t$ as well as $f\sigma_8(Z)$ against the simplest disformal framework (with $\alpha =0.08, \beta = 0.0$ and $D_m = 1.0$) is found to be $\leq 0.5 \%$ and $\leq 0.35 \%$ respectively at $Z=0$, for any value of $\beta$ in the range of $0< \beta<1000$. 

\begin{figure}[h!] 
\vspace{-0.2cm}
\center
\includegraphics[scale=0.375]{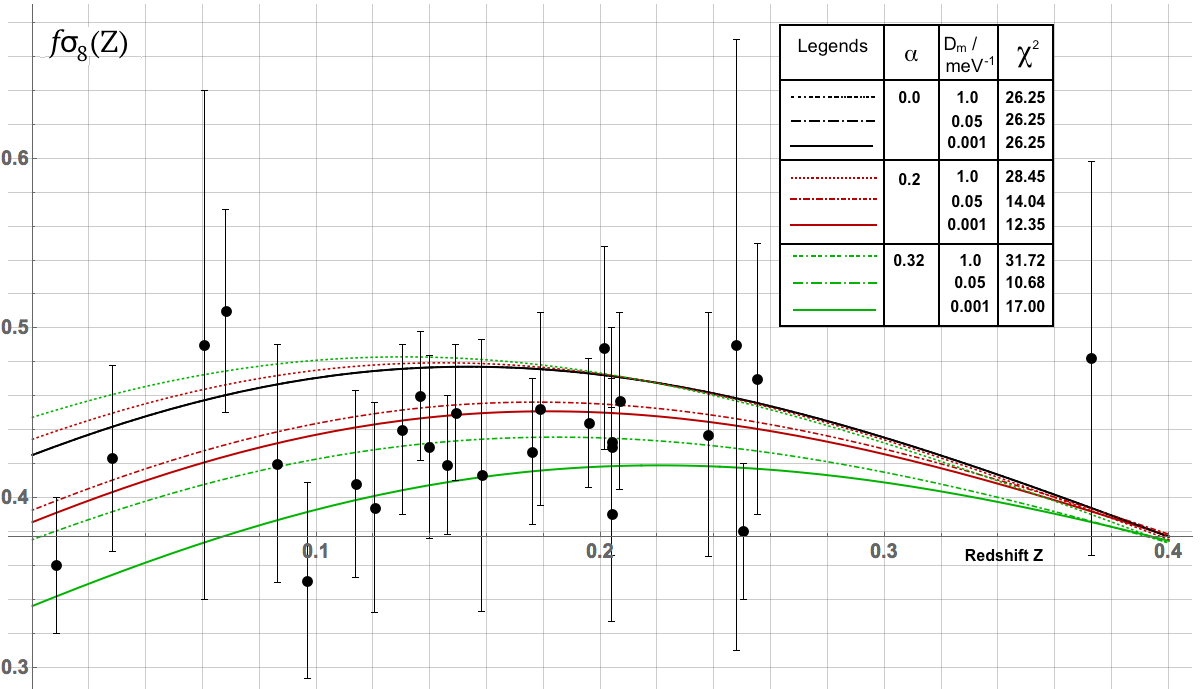} 
\caption{\label{fig2}The effect of quintessential mass on the numerical $f\sigma_8(Z)$ curve of the disformally coupled quintessence model is shown, together with RSD data \cite{delaTorre2013, Okada2015,Guzzo2008,Chuang2013, Blake2012,SDSS2006, GilMarin2015,Tojeiro2012, BOSS2016,Samushia2012,Blake2011, 2dFGRS2004,Howlett2014,Beutler2012,Song2008}.  The black, red and green lines correspond to  $\alpha = 0.0,\; 0.2$ and $0.32$ respectively. For each color line, the solid, dot-dashed and dotted lines correspond to $D_m = 0.001,\; 0.05,\;1.0\; \rm{meV}^{-1}$ respectively. All black lines coincides with each other. Other parameters are set as $\sigma^0_8=0.818$  from Ref. \cite{Barros2018} and $\beta$ as unity.}
\end{figure}

The Figure \ref{fig1} and Figure \ref{fig2}  show the effect of the quintessential mass on the evolution of the total matter perturbations $\delta_t $, and the numerical $f\sigma_8(Z)$ curve respectively, whereby the parameters $\sigma^0_8$ and $\beta$ values are set to be $\sigma^0_8=0.818$ and $\beta=1.0$. The following remarks are made for both Figure \ref{fig1} and Figure \ref{fig2}:
\begin{enumerate}
\item When the conformal parameter  $\alpha$ is set to zero, the DCQ model approaches towards $\Lambda$CDM model in the limit $D_m \rightarrow 0$ (i.e., $M \rightarrow \infty$).
\item For non-zero values of $\alpha$ and $\beta$, the conformal nature of the model dominate at lower values of $D_m$ (i.e., higher mass $M$) and disformal framework becomes more significant at higher values of $D_m$ (i.e., lower mass $M$) as shown in Ref. \cite{Dusoye2020}.
\item Hence with $\beta$ set to unity and small $D_m$, the model tend to behaves conformally, similar to the results in Ref. \cite{Barros2018}.
\item For a fixed choice of $\alpha$, the parameter $D_m$ induces a gradual disformal effect. When $D_m$ is very small (i.e., higher mass $M$), $f\sigma_8(Z)$ curve coincides with the purely conformal case, because the disformal coefficient $D(\phi)$ becomes zero, even if $\beta$ is non-zero. As $D_m$ is increased (i.e., decreasing mass $M$), a disformal effect is induced such that the rate of evolution of $\delta_t $ then increases leading to higher $f\sigma_8(0)$ value. The structures cluster faster when compared to the respective conformal case. 
\end{enumerate}


A Bayesian analysis \cite{William1992} of this DCQ model was carried out to the find the set of the parameters, which will fit the RSD data \cite{delaTorre2013, Okada2015,Guzzo2008,Chuang2013, Blake2012,SDSS2006, GilMarin2015,Tojeiro2012, BOSS2016,Samushia2012,Blake2011, 2dFGRS2004,Howlett2014,Beutler2012,Song2008}, in a similar manner that was done in Ref. \cite{Barros2018}. In our analysis, the likelihood $ \mathit{L}(\beta)$ is found to be equally probable for all values $\beta$ and this re-confirms that the parameter $\beta$ does not affect the evolution of the total matter perturbation. For this reason, the disformal parameter $\beta$ is fixed to unity, as an arbitrary choice. The set of parameters $\Theta_{\mu}$ is then reduced to $\Theta_{\mu} \in  \left\lbrace \sigma_0,\;\alpha,\;D_m \right\rbrace$ only. 

\begin{figure}[h!] 
\center
\includegraphics[scale=0.4]{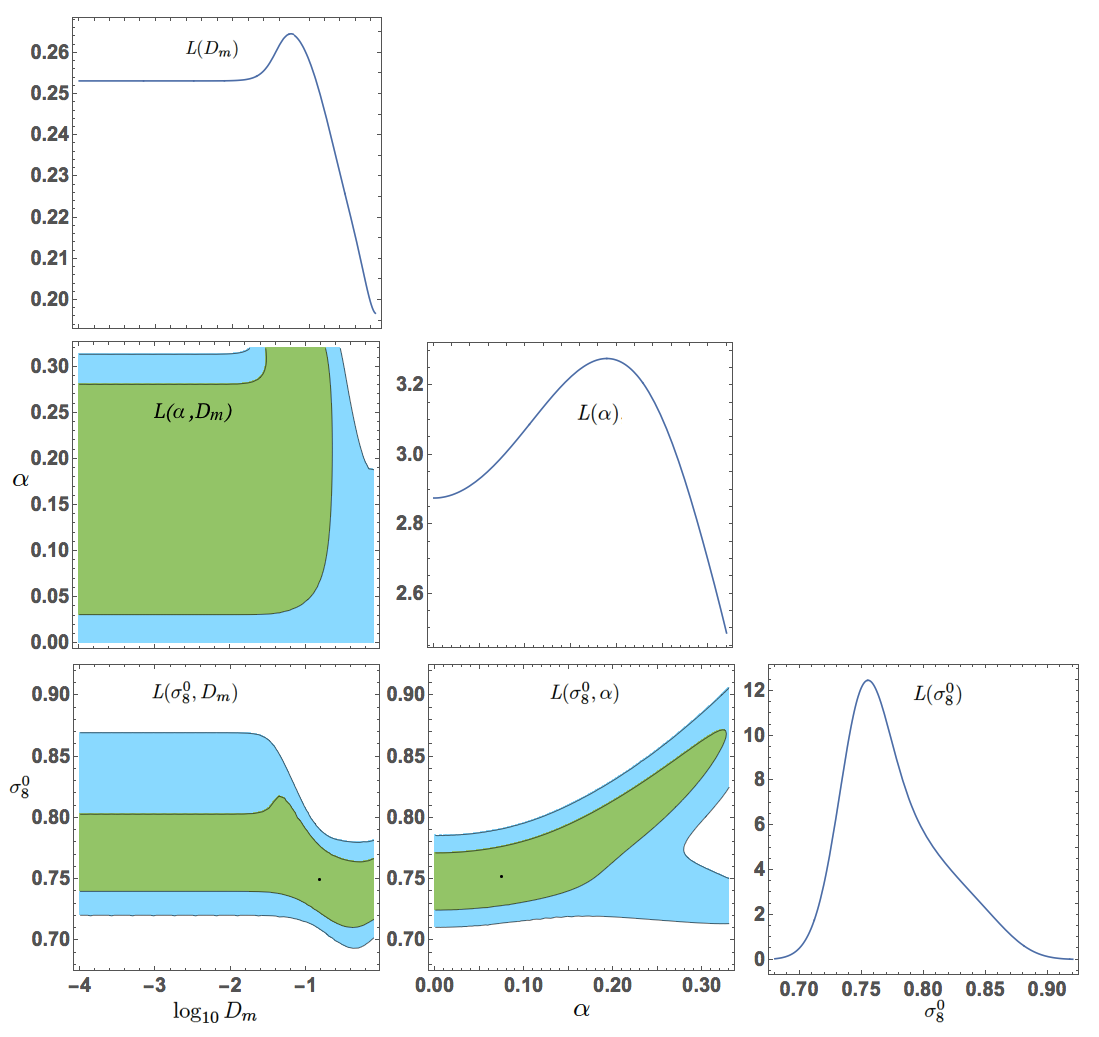} 
\caption{\label{fig3} The marginalised surface Likelihoods $\mathit{L}(\alpha,D_m)$, $\mathit{L}(\sigma^0_8,D_m)$ and $\mathit{L}(\sigma^0_8, \alpha)$ are shown in the top view, where the green and the light blue regions indicate with $68 \%$ and $95 \%$ confidence as to where the correct set of parameters $\Theta_{\mu}$ exists within each respective 2D parameters space. The marginalised likelihood curves $\mathit{L}(D_m)$, $\mathit{L}(\alpha)$, and $\mathit{L}(\sigma^0_8)$ are also shown.}
\end{figure}

A three-dimensional grid evaluation was carried out with $n=83, 000$ points to find the most probable set of parameters, which most agree with the RSD data \cite{delaTorre2013, Okada2015,Guzzo2008,Chuang2013, Blake2012,SDSS2006, GilMarin2015,Tojeiro2012, BOSS2016,Samushia2012,Blake2011, 2dFGRS2004,Howlett2014,Beutler2012,Song2008}. The bounds of the 3D parameters space are set as (i)  $0.65 \;\leq\; \sigma^0_8 \;\leq\; 0.95$, (ii) $0 \;\leq\; \alpha \;\leq\; 0.35 $, and (iii) $10^{-4} \,\leq\, D_m \,\leq\; 1.0 $. The best fit set of parameters is found to be (within $68\%$ Confidence Level) :
\vspace{-0.5cm}
 \begin{eqnarray}\label{E4.47}
\alpha\vert_{\rm{BF}} = 0.32\pm 0.09, \hspace{0.5cm}
\sigma^0_8\vert_{\rm{BF}} = 0.81\pm 0.08,  \hspace{0.5cm}
 D_{m\vert\rm{BF}} = 0.056^{+0.217}_{-0.056}\;.   \hspace{0.5cm}
\end{eqnarray}
We find that our obtained $\sigma^0_8\vert_{\rm{BF}}$  value (within $68\%$ Confidence Level) seems to be consistent with the Planck (2018) value of $\sigma^0_8 = 0.811 \pm 0.006$ \citep{Plank2018}. The best fit $\sigma^0_8\vert_{\rm{BF}}$ corresponds to a $\sigma^0_8\vert_{\rm{BF}}$ value, which is close to the conformal constraints from Ref. \citep{Barros2018}. The Hubble tension is observed because the error margin of the DES's constraints $ \sigma^0_8 = 0.762^{+0.025}_{-0.024}$ \cite{Joudaki2019} do not overlap the error margin of the Plank (2018)'s constraints. However, the $2.5\sigma$ Hubble tension seems to be relaxed by the predicted $\sigma^0_8$ values from CCQ and DCQ model, because their error margin are wide enough to overlap both constraint values by Planck and DES data.\\

Figure \ref{fig3} shows the marginalisation of the main Likelihood $\mathit{L}(\Theta_{\mu})$ over one parameter to yield the surface likelihoods $\mathit{L}(\alpha,D_m)$, $\mathit{L}(\sigma^0_8,D_m)$ and $\mathit{L}(\sigma^0_8, \alpha)$. These surface likelihoods are shown in the top view, where the green and the light blue regions indicate $68 \%$ and $95 \%$ confidence. Figure \ref{fig3} also shows the marginalisation of the main Likelihood $\mathit{L}(\Theta_{\mu})$ over two parameters to yield the 1D likelihood curves denoted by $\mathit{L}(D_m)$, $\mathit{L}(\alpha)$, and $\mathit{L}(\sigma^0_8)$. The peak value of the surface Likelihood $\mathit{L}(\sigma^0_8,D_m)$ corresponds to ($\alpha = 0.074$ and $\sigma^0_8 = 0.75$), which agrees with the conformal constraints, previously obtained in Ref. \citep{Barros2018}. For range $0 <\alpha < 0.15$, the surface likelihood $\mathit{L}(\sigma^0_8, \alpha)$ maintains the characteristics of the conformal constraints with RSD data (See Figure 4 in Ref. \citep{Barros2018}). The 1D likelihood curves $\mathit{L}(\alpha)$ and $\mathit{L}(\sigma^0_8)$ also manifest this reasoning. Any deviation from Figure 4 in Ref. \citep{Barros2018} is attributed to the disformal dependence of this quintessence model. In the limit $D_m \rightarrow 0$, i.e., when the model starts to behave conformally, there can be various values of $D_m$ for a given value of $\sigma^0_8$ in the surface Likelihood $\mathit{L}(\sigma^0_8,D_m)$. This appears similarly in the surface likelihood $\mathit{L}(\alpha, D_m)$ for a given value of $\alpha$. This could suggest that the parameter $D_m$ is actually degenerate in that limit.\\

\vspace{-0.5cm}
\begin{table}[ht]
\center
\begin{tabular}{|  l ||  c  c   c   c | c | c  |c|}
 \hline
 & &&  &&  & & \\ [0.01ex]
Model & $\sigma^0_8\vert_{\rm{BF}}$ &$\alpha\vert_{\rm{BF}}$ &$ D_{m\vert\rm{BF}}$ & $\beta$ & N$_{\rm{f}}$ & $\chi^2_{\rm{min}}$& $\chi^2_{\rm{red}}$\\[1.5ex]
 \hline  \hline
  & &&  &&  & & \\ [0.01ex]
$\Lambda$CDM & $0.750 \pm 0.024$ &$0$ & $0$ &$0$ & $1$ & $11.4413$& $0.4400$\\ [1.5ex]
 \hline 
& &&  &&  & & \\ [0.01ex]
CCQ & $0.818^{+0.115}_{-0.088}$ &$0.079^{+0.059}_{-0.067}$ & $0$ &$0$ & $2$ & $11.0946$& $0.4438$\\ [1.5ex]
 \hline 
 & &&  &&  & & \\ [0.01ex]
 DCQ & $0.81\pm0.08$ &$0.32\pm0.09$ & $0.056^{+0.217}_{-0.056}$ & $1.0$& $4$ & $10.664$& $0.463$\\ [1.5ex]
 \hline 
\end{tabular}
\caption{\label{T4.1}The Bayesian analyses provided the reduced Chi-square estimations $\chi^2_{\rm{red}}$ for the best-fit parameters  for the $\Lambda$CDM model, the CCQ model of Ref. \citep{Barros2018}, and our studied DCQ model in $\Lambda$CDM background. N$_{\rm{f}}$ is number of fitted parameters.}
\end{table}

The reduced Chi-square estimations of the best-fit parameters $\chi^2_{\rm{red}}$ as well as the number of fitted parameter N$_{\rm{f}}$ for the $\Lambda$CDM model, the CCQ model of section and our studied DCQ model in $\Lambda$CDM background are summarised in Table \ref{T4.1}. Compared to $\Lambda$CDM and the CCQ models, the minimum Chi-square $\chi^2_{\rm{min}}$ of the DCQ model is found to be lower but its reduced Chi-square $\chi^2_{\rm{red}}$ is in fact larger. This is counter-intuitive but could be accounted to the degeneracies of the parameters of the DCQ model, which are seen in the following: (i) the negligible relative difference of $f\sigma_8(Z)$ curve due to the effect of $\beta$, (ii) equi-probability of $L(\beta)$ for large range of $\beta$, (iii) the marginalised surface likelihoods appears as bands instead of Gaussian ellipses because within certain region of the parameter spaces, the surface likelihoods are actually equi-probable. The relatively higher value of reduced Chi-square for the DCQ model could also be due to large uncertainties associated with the values of the best fit parameters.\\


From these facts, one cannot really state whether the DCQ model is better than $\Lambda$CDM or the CCQ model, until these degeneracies are clarified. The degeneracies may not necessarily imply that the studied disformal formalism is an incorrect model. In our RSD data \cite{delaTorre2013, Okada2015,Guzzo2008,Chuang2013, Blake2012,SDSS2006, GilMarin2015,Tojeiro2012, BOSS2016,Samushia2012,Blake2011, 2dFGRS2004,Howlett2014,Beutler2012,Song2008}, we deal only with $f\sigma_8$ values and redshift $Z$. However, the Ref. \cite{Perenon2019} mentions that the combination of separate data of growth $f$ and matter fluctuation amplitude $\sigma^0_8$ is capable of actually breaking degeneracies, which the quantity $f\sigma_8$ would not be able to do intuitively.  As future work, we intend to elucidate the degeneracies of the DCQ model using the combination of ($f + \sigma^0_8$) data, with the same methodology used in Ref. \cite{Perenon2019}. Moreover, it is also possible that the use of the late-time RSD data is indicating only one aspect of the big picture in understanding the Hubble tension. Another way to further investigate and to reduce the degenerate variables is for us to consider constraining the studied DCQ model with CMB data or BAO data or Supernovae surveys.\\

\section{Conclusion}
In conclusion, we have explored the perturbations of the disformally coupled quintessence,as a replacement for the cosmological constant, while it mimics the $\Lambda$CDM background. The DCQ model is analysed numerically by solving the system of  evolution equations \eqref{E3.15} and \eqref{E3.16} to obtain $\delta_c$ and $\delta_b $ in terms of the cosmological redshift $Z$. Using the definition \eqref{E4.3}, we plot the numerical $f\sigma_8(Z)$ curve for several sets of parameters. The effect of the quintessential mass is seen in the evolution of total matter perturbation (See Figure \ref{fig1}) and the numerical $f\sigma_8(Z)$ curve (See Figure \ref{fig2}). For a given choice of $\alpha$, when a decreasing quintessential mass is used for the model, a disformal effect seems to be induced. This disformal effect leads to a faster rate of evolution for total matter density contrast and a higher $f\sigma_8(0)$ value, which imply that the  large-scale structures cluster faster  compared to the its respective conformal case.

 The free parameters of the DCQ model (i.e., $ \sigma^0_8,\;\alpha,\;\beta ,\; D_m $) are investigated with Bayesian statistics using the RSD data \cite{delaTorre2013, Okada2015,Guzzo2008,Chuang2013, Blake2012,SDSS2006, GilMarin2015,Tojeiro2012, BOSS2016,Samushia2012,Blake2011, 2dFGRS2004,Howlett2014,Beutler2012,Song2008}. The likelihood $ \mathit{L}(\beta)$ is found to be constant for all values of $\beta$ and confirms that $\beta$ does not affect the evolution of the total matter perturbation. The Figure \ref{fig3} shows the marginalised surface likelihoods $\mathit{L}(\alpha,D_m)$, $\mathit{L}(\sigma^0_8,D_m)$ and $\mathit{L}(\sigma^0_8, \alpha)$ in topview, where the green and light blue indicate the $68 \%$ and $95 \%$ confidence regions. The best fit set of parameters is found to be $\sigma^0_8\vert_{\rm{BF}} = 0.81\pm 0.08$, $\alpha\vert_{\rm{BF}} = 0.32\pm 0.09$ and $D_{m\vert\rm{BF}} =0.056^{+0.217}_{-0.056}$ within $68 \%$ confidence level, at the minimum chi-square value $\chi^2_{\rm{min}} = 10.664$. The best fit value of $\sigma^0_8\vert_{\rm{BF}}$ is consistent with latest Planck (2018) value of $\sigma^0_8 = 0.811 \pm 0.006$ \citep{Plank2018}. The existent Hubble tension seems to be relaxed by the predicted $\sigma^0_8\vert_{\rm{BF}}$ from the studied DCQ model.The minimum Chi-square $\chi^2_{\rm{min}}$ of the DCQ model is found to be lower but its reduced Chi-square $\chi^2_{\rm{red}}$ is in fact larger than $\Lambda$CDM and the CCQ models,(see Table \ref{T4.1}). This is counter-intuitive but might be explained by the degeneracies of the parameters of the DCQ model. Therefore, one cannot really state whether the DCQ model is better than $\Lambda$CDM or the CCQ model, until these degeneracies are elucidated. As future work, we shall clarify the degenerate variables of the DCQ model using the combination of ($f + \sigma^0_8$) data (with same statistical techniques from Ref. \cite{Perenon2019}). As future endeavours, we shall allow the parameters $\alpha$ and $\beta$ to influence the entire evolution of density perturbations from the initial redshift $Z_i$ and find consistent constraints on DCQ model by fitting directly the multipoles of the power spectrum since the density and the velocity power spectra is expected to deviate from GR between the initial and observed redshift. Lastly, it would also be worthwhile to constrain the studied DCQ model with Planck CMB data and to investigate the disformal effects on the early universe.

\section{Appendices}
\hspace{0.7cm}The Ref. \cite{Gleyzes2015a} has already derived the second-order perturbation theory of a scalar field which is disformally coupled with CDM in a Horndesksi model, defined in terms of $\alpha$ basis (in the unitary gauge). Although the derivations in Ref. \cite{Gleyzes2015a} are more general and apply to many models, we shall summarise the steps to obtain the  perturbation equations for specifically our DCQ model in terms of dynamical variables \eqref{EDV} for the sake of completeness.

\subsection{Second-order perturbation equations of DCQ model}\label{A1}
The equations \eqref{E3.7}-\eqref{E3.8} are first simplified with choice of $w_{\rm{f}}= \dot{w}_f = \nabla^2\pi_{\rm{f}} = 0$. The time-derivative of equation \eqref{E3.7} is then carried out and the equation \eqref{E3.8} is used to substitute for $\dot{\theta}_{c}$.  The perturbed Einstein equation \eqref{E3.5}, with $\textit{c}_s^2 =0$, can used to replace $\ddot{\Psi}$ (e.g., as done in Ref \cite{Gleyzes2015a}). Furthermore, we assume the approximation of $\textit{k}^2\Psi \gg a^2H^2\Psi$ and $\textit{k}^2\chi \gg a^2H^2\chi$, i.e., the Newtonian limit for sub-Hubble scales. The perturbed Einstein equation \eqref{E3.4} and  the perturbed Klein-Gordon \eqref{E3.6} can be similarly approximated in the Newtonian limit \cite{Gleyzes2015a}, to render the expressions below: 
\begin{equation}\label{E6.1} 
 \dfrac{\nabla^2\Psi}{a^{2}} \approx \dfrac{3}{2} H^2(\Omega_{c}\delta_{c} + \Omega_{\mathtt{B}}\delta_{\mathtt{B}})
\hspace{0.1cm}\text{,}\hspace{0.5cm} \dfrac{\nabla^2\chi}{a^{2}} \approx -Q_1 
\hspace{0.5cm}\text{and}\hspace{0.5cm} \dfrac{\nabla^2\dot{\chi}}{a^{2}} \approx -(\dot{Q}_1+2HQ_1)\;.
\end{equation}
The $\nabla^2\dot{\chi}$ can also be computed, which is useful for later derivations. The terms $\nabla^2\Psi$ and $\nabla^2\chi$, which appears in the obtained evolution equation and which originally comes from equation \eqref{E3.8}, are then substituted by the expressions \eqref{E6.1}. The second-order evolution equation \eqref{E3.16} is then obtained after re-expressing all its functions and derivatives in terms of the dynamical variables \eqref{EDV} as well as the normalisation \eqref{E3.17}, and is then simplified with conservation equation \eqref{E2.15} to replace $(x^{\prime} +\hat{H}x)$.

\subsection{The normalised derivative of the interaction term}\label{A2}
The normalised derivative of the interaction term  \eqref{E3.18}, i.e. $\hat{q}_0$, which is required in equation \eqref{E3.16} can be obtained as follows for arbitrary choice of $\lambda_C$ and $\lambda_D$. The interaction term $Q(\phi)$ for a single coupled fluid can be computed by evaluating the covariant derivative of time components for the energy-momentum tensor \eqref{E2.5} and expressed in terms of the dynamical variables \eqref{EDV}
\vspace{-0.2cm}
\begin{eqnarray} \label{E6.2}
\left(1-3\sigma x^2\right)\hspace{-0.2cm}\,\left(\dfrac{\kappa Q_0}{3H^2 z^2}\right)&=&\dfrac{\lambda_C}{2}(1-3w_{c})+3\sqrt{3}w_{c}\sigma x -\sqrt{3}\dfrac{\sigma }{H}\left(\dot{x} + \dfrac{\dot{H}}{H}x \right)   \nonumber \\
&&  + \dfrac{3}{2}\left(\lambda_D -2\lambda_C\right)\sigma x^2. 
\end{eqnarray}
The EOS for coupled DM is obtained by setting $w_{c}=0$. When the equation \eqref{E6.2} is expressed in terms of e-fold $N$ and the conservation equation \eqref{E2.15} is used to replace $(x^{\prime} +\hat{H}x)$, one obtains the interaction term \eqref{EQ0} as expected. The derivative of the equation \eqref{E6.2} with respect to cosmic time is carried out and is expressed in terms of e-fold $N$ to yield:
\begin{eqnarray} \label{E6.3}
&&\left(1-3\sigma x^2\right)\hspace{-0.2cm}\,\left(\dfrac{\kappa Q_0^{\prime}}{3H^2 z^2}\right) = 2\hspace{-0.1cm}\left(1-3\sigma x^2\right)\hspace{-0.2cm}\,\left(\dfrac{\kappa Q_0}{3H^2 z^2}\right)\hspace{-0.2cm}\left(\dfrac{H^{\prime}}{H}+ \dfrac{z^{\prime}}{z}\right) +\dfrac{\kappa Q_0}{H^2 z^2}\left(\sigma^{\prime} x^2 + 2\sigma x^{\prime}x\right)  \nonumber \\ 
&&+\dfrac{\lambda_C^{\prime}}{2} + \dfrac{3}{2}(\lambda_D^{\prime} -2\lambda_C^{\prime})\sigma x^2+\dfrac{3}{2}(\lambda_D -2\lambda_C)\hspace{-0.15cm}\left(\sigma^{\prime}x^2 + 2\sigma xx^{\prime}\right) - \sqrt{3}\left(\sigma^{\prime}-\dfrac{H^{\prime}}{H}\sigma \right)\hspace{-0.2cm}\left(x^{\prime} + \dfrac{H^{\prime}}{H}x \right) \nonumber\\
&& -\sqrt{3}\sigma\hspace{-0.15cm}\left(x^{\prime\prime} + 2\dfrac{H^{\prime}}{H}x^{\prime} +  \dfrac{H^{\prime\prime}}{H}x \right)\;.
\end{eqnarray}
The conservation equations \eqref{E2.15} and \eqref{E2.16}  are inserted in to substitute  $(x^{\prime} +\hat{H}x)$ and $z^{\prime}/z$ respectively. The derivative of Raychaudhuri equation in \eqref{E2.19} is used to replace $H^{\prime\prime}/H$. Some of the terms get simplified with the normalised variables \eqref{E3.17} and using $\lambda_C^{\prime}=\lambda_D^{\prime} =0$ as per choices in equations \eqref{E2.20} and \eqref{E2.22} and then, the normalised derivative $\hat{q}_0 $ is hence obtained as in equation \eqref{E3.18}.

\subsection{The normalised derivative of the perturbed interaction}\label{A3}
The normalised derivative of the perturbed interaction, i.e. $\hat{q}_1$, which completes the equation \eqref{E3.16} can be obtained as follows for arbitrary choice of $\lambda_C$ and $\lambda_D$. The perturbed interaction term $Q_1$, which is given by \eqref{E3.9}-\eqref{E3.14}, is re-expressed in terms of the dynamical variables \eqref{EDV} in cosmic time as below: 
\begin{eqnarray}\label{E6.4}
\kappa Q_1 &=& \dfrac{-3H^2z^2}{\left(1-3\sigma x^2 + 3\sigma z^2\right)}\left(\mathcal{A}_1\delta_{c}  + \mathcal{A}_2\dot{\Phi}   +\mathcal{A}_3\Psi +  \mathcal{A}_4\dot{\chi} + \mathcal{A}_5\chi \right) \hspace{3.5cm}\\
\text{with} \hspace{1.0cm}\nonumber \\ \label{E6.5}
\mathcal{A}_1 &=& - \dfrac{\lambda_C}{2} -\dfrac{\sqrt{3}\sigma}{H}\left(\dot{x}+ \dfrac{\dot{H}}{H}x \right) -\dfrac{3\sigma x^2}{2}\left(\lambda_D - 2\lambda_C \right)\;, \\ 
 \label{E6.6}
\mathcal{A}_2 &=& \dfrac{3\sqrt{3}}{H}(1+w_{c})\sigma\textit{x} \;, \\
 \label{E6.7}
\mathcal{A}_3 &=& 6\sqrt{3}(1+w_{c})\sigma \textit{x} + 3\sigma\textit{x}^2\left(\lambda_D - 2\lambda_C + 2\dfrac{\kappa Q_0}{3H^2 z^2}\right) \;, \\
\label{E6.8}
\mathcal{A}_4 &=& -3\dfrac{\kappa\sigma}{H}(1+w_{c}) -\sqrt{3}\dfrac{\kappa\sigma}{H}\textit{x}\left(\lambda_D - 2\lambda_C + 2\dfrac{\kappa Q_0}{3H^2 z^2}\right) \;,
\end{eqnarray} 
\vspace{-0.5cm}
\begin{eqnarray}
\mathcal{A}_5 &=& \dfrac{\kappa \Lambda_C}{2} - \dfrac{\kappa\sigma}{a^2H^2}(1+w_{c})\nabla^2 - \dfrac{\kappa\sigma}{H^2}V_{,\phi\phi} + \lambda_D \dfrac{\kappa^2\sigma}{H^2}V_{,\phi} - 3\sqrt{3}\kappa \sigma \lambda_D \textit{x}\, (1+w_{c}) \nonumber \\
&& -3\kappa\sigma x^2\left(\Lambda_C + \lambda_C^2 + \lambda_C\lambda_D +  \dfrac{\Lambda_D}{2}\right) - \kappa\left[\lambda_C + 3\lambda_D\sigma\left(z^2-x^2\right)\right]\dfrac{\kappa Q_0}{3H^2 z^2}\;, \hspace{1.0cm}\label{E6.9}
\end{eqnarray} 
such that the functions $\mathcal{A}_m $ are re-normalised as $\mathcal{A}_m \equiv \mathcal{B}_m/\kappa C$ with $m = 1,\dots,5$. The $\nabla^2$ operator appearing in the second term of the coefficient $\mathcal{A}_5$ in \eqref{E6.9} is acting upon the variable $\chi$ in equation \eqref{E6.4}.  The relation $\Psi = \Phi$ was applied to equation \eqref{E6.4} and the EOS was set to $w_{c} =0$ since the coupled fluid is DM in this model. Under the approximation of the Newtonian limit, the perturbed interaction term \eqref{E6.4} then reduces to the following.
\begin{eqnarray}\label{E6.10}
\kappa Q_1\left(1-3\sigma x^2+ 3\sigma z^2\right) \approx -3H^2z^2\left[\mathcal{A}_1\delta_c  +  \mathcal{A}_5\chi \right] \approx -3H^2z^2\left[\mathcal{A}_1\delta_c  -  \dfrac{\kappa\sigma}{H^2}\left(\dfrac{\nabla^2\chi}{a^{2}}\right)\right]\;,\hspace{0.5cm}
\end{eqnarray}
where only the second term of the coefficient $\mathcal{A}_5$ in equation \eqref{E6.9} survives because it is of same order of $\textit{k}^2\chi $. After having inserted  $a^{-2}\nabla^2\chi\approx -Q_1 $, i.e., from the approximation \eqref{E6.1} of the pertubed Klein Gordon equation, and after having substituted for the normalised variable $\tilde{Q}_1$ (as defined in equation \eqref{E3.17}), one can simplify to obtain:
\vspace{-0.2cm}
\begin{eqnarray}\label{E6.11}
\tilde{Q}_1 \approx - \dfrac{\mathcal{A}_1\delta_c }{1-3\sigma x^2+6\sigma z^2}\;.
\end{eqnarray}
The derivative of perturbed interaction \eqref{E6.4}  with respect to cosmic time is computed:
\begin{eqnarray} \label{E6.12}
&&\kappa \dot{Q}_1\left(1-3\sigma x^2+ 3\sigma z^2\right)  - \kappa Q_1\left(3\dot{\sigma}x^2 + 6\sigma x\dot{x} -3\dot{\sigma}z^2 - 6\sigma z\dot{z} \right) = -6(H\dot{H}z^2
\hspace{1.5cm} \nonumber  \\
&&+\; H^2z\dot{z})\left[\mathcal{A}_1\delta_{c}  +  \mathcal{A}_2\dot{\Phi} + \mathcal{A}_3\Psi + \mathcal{A}_4\dot{\chi} + \mathcal{A}_5\chi \right] - 3H^2z^2\left[\dot{\mathcal{A}}_1\delta_{c} \mathcal{A}_1\dot{\delta}_{c}  + \dot{\mathcal{A}}_2\dot{\Phi} \right. \nonumber\\
&&+\left. \mathcal{A}_2\ddot{\Phi}  +\dot{\mathcal{A}}_3\Psi +\mathcal{A}_3\dot{\Psi} +  \dot{\mathcal{A}}_4\dot{\chi} + \mathcal{A}_4\ddot{\chi} + \dot{\mathcal{A}}_5\chi + \mathcal{A}_5\dot{\chi}   \right]\;.
\end{eqnarray}
Applying the relation $\Psi = \Phi$, the equation \eqref{E6.12} is reduced under the Newtonian limit.
\begin{eqnarray} \label{E6.13}
&&\kappa \dot{Q}_1\left(1-3\sigma x^2+ 3\sigma z^2\right) = \kappa Q_1\left(3\dot{\sigma}x^2 + 6\sigma x\dot{x} -3\dot{\sigma}z^2 - 6\sigma z\dot{z} \right) -6(H\dot{H}z^2 + H^2z\dot{z})\mathcal{A}_1\delta_{c}  \nonumber \\
&&- 6(H\dot{H}z^2 + H^2z\dot{z})\mathcal{A}_5\chi   - 3H^2z^2\left(\dot{\mathcal{A}}_1\delta_{c} + \mathcal{A}_1\dot{\delta}_{c}\right) - 3H^2z^2\dot{\mathcal{A}}_5\chi - 3H^2z^2\mathcal{A}_5\dot{\chi}\;, 
\end{eqnarray}
where only the second term of the coefficient $\mathcal{A}_5$ as in equation  \eqref{E6.9} and its derivatives can survive because they are of same order of $\textit{k}^2\chi $. The following four terms in equation \eqref{E6.13} are approximated under the Newtonian limit and  re-expressed using the relations \eqref{E6.1} and become as below:
\begin{small}
\begin{eqnarray} \label{E6.14}
6H\dot{H}z^2\mathcal{A}_5\chi &\approx & \;   -6\kappa\sigma z^2 \dfrac{\dot{H}}{H} \left(\dfrac{\nabla^2\chi}{a^{2}}\right) \approx + 6\dfrac{\dot{H}}{H}\sigma z^2\kappa Q_1 \;,\\
6H^2z\dot{z}\mathcal{A}_5\chi  &\approx&  -6z\dot{z}\kappa\sigma\left(\dfrac{\nabla^2\chi}{a^{2}}\right) \approx + 6z\dot{z}\sigma\kappa Q_1 \;,\nonumber \\
3H^2z^2 \mathcal{A}_5\dot{\chi}   &\approx & -3z^2\kappa\sigma\left(\dfrac{\nabla^2\dot{\chi}}{a^{2}}\right) \approx + 3z^2\sigma\kappa(\dot{Q}_1 + 2HQ_1)\;,\nonumber\\
3H^2z^2 \dot{\mathcal{A}}_5\chi &\approx & -3z^2\kappa\left[\dot{\sigma} - \left( 2H + 2 \dfrac{\dot{H}}{H}\right)\sigma \right]\hspace{-0.2cm}\left(\dfrac{\nabla^2\chi}{a^{2}}\right)\;, \nonumber
\approx  +3z^2\left[\dot{\sigma} - \left( 2H + 2 \dfrac{\dot{H}}{H}\right)\sigma \right]\kappa Q_1\;. \nonumber 
\end{eqnarray}
\end{small}
Those four approximated terms  \eqref{E6.14} are inserted into equation \eqref{E6.13}. By dividing by $3H^2z^2$ and re-arranging the $\dot{Q}_1$ to one side, one can simplify equation \eqref{E6.13} due to some cancelling terms. The simplified form of equation \eqref{E6.13} is re-expressed in terms of $N$.
\begin{eqnarray}\label{E6.15}
\dfrac{\kappa Q_1^{\prime}}{3H^2z^2}\left(1-3\sigma x^2+ 6\sigma z^2\right)  &=& \dfrac{\kappa Q_1}{3H^2z^2}\left(3\sigma^{\prime}x^2 + 6\sigma xx^{\prime} -6\sigma^{\prime}z^2 - 12\sigma zz^{\prime} \right)  \nonumber\\
&& -2\left(\dfrac{H^{\prime}}{H} + \dfrac{z^{\prime}}{z} \right)\mathcal{A}_1\delta_{c}
 - \mathcal{A}_1^{\prime}\delta_{c} - \mathcal{A}_1\delta_c^{\prime}\;.
\end{eqnarray}
The conservation equation \eqref{E4.3} is used to substitute $z^{\prime}/z$ and the normalised variables \eqref{E3.17} are inserted into equation \eqref{E6.15}. The derivative $\hat{q}_1$ as per equation \eqref{E3.19}-\eqref{E3.20} is thus obtained after some arrangement of equation \eqref{E6.15}.

\section{Acknowledgements}
AdlCD and AD acknowledge financial support from NRF Grants No.120390, Reference: BSFP190416431035, and No.120396, Reference: CSRP190405427545, and No 101775, Reference: SFH150727131568. AdlCD also acknowledges financial support from Project COOPB 204064, I-COOP+2019 CSIC Actions from the Spanish Ministry of Science, MICINN Project No. PID2019-108655GB-I00 from the European Regional Development Fund and Spanish Research Agency (AEI), and support from Projects Nos. CA15117 and CA16104 from COST Action EU Framework Programme Horizon 2020. AdlCD thanks the hospitality of the Institute of Theoretical Astrophysics - University of Oslo (Norway) during the later steps of the manuscript. AdlCD and NN also acknowledge funding from the University of Cape Town Visiting Scholars Fund 2018. 
This research was supported by Funda\c{c}\~ao para a Ci\^encia e a Tecnologia (FCT) through the research grants: UIDB/04434/2020, UIDP/04434/2020, PTDC/FIS-OUT/29048/2017 (DarkRipple),  COMPETE2020: POCI-01-0145-FEDER-028987 \& FCT: PTDC/FIS-AST/28987/2017 (CosmoESPRESSO), CERN/ FIS-PAR/0037/2019 (MGiCAP) and IF/00852/2015 (Dark Couplings). PKSD thanks First Rand Bank for financial support.


\noindent\rule{16cm}{0.4pt}\newpage
\end{document}